\documentclass[reprint,amsmath,amssymb, aps,floatfix,longbibliography,]{revtex4-2}
\usepackage{graphicx}
\usepackage{float}
\usepackage{dcolumn}
\usepackage{bm}

\usepackage[pdf]{pstricks}
\usepackage{pst-all}
\usepackage{epsfig}
\usepackage{epstopdf}
\usepackage{mathtools}
\usepackage{amsmath}

\DeclarePairedDelimiter\abs{\lvert}{\rvert}%
\DeclarePairedDelimiter\norm{\lVert}{\rVert}

\makeatletter
\let\oldabs\abs
\def\abs{\@ifstar{\oldabs}{\oldabs*}}
\let\oldnorm\norm
\def\norm{\@ifstar{\oldnorm}{\oldnorm*}}
\makeatother

\PrependGraphicsExtensions{.tif}

\begin{document}
\setlength{\abovedisplayskip}{3pt}
\setlength{\belowdisplayskip}{3pt}
\preprint{APS/123-QED}

\title{Radially Locked Sun-Ray Patterns in \\Autocatalytic Reaction-Diffusion-Advection Systems}
\author{S. N. Maharana, L. Negrojevi\'{c}, A. Comolli and A. De Wit}%
\affiliation{Nonlinear Physical Chemistry Unit, Universit\'e libre de Bruxelles (ULB), 1050 Brussels, Belgium}%
\date{\today}

\begin{abstract}
Traveling fronts ubiquitous in physics, chemistry, and biology are prone to transverse cellular deformations due to diffusive or convective instabilities. Here we show both theoretically and experimentally that new patterns can be obtained if the destabilization is triggered around a front locked radially by advection. Specifically, angularly shifting sun-ray-like patterns can develop around radially advected autocatalytic fronts due to a diffusive instability developing when the autocatalyst  X and the reactant Y diffuse at different rates. The properties of these shining-star structures can be controlled by tuning the flow rate $Q$ and the ratio of diffusion coefficients $\delta$ as evidenced by linear stability analysis, nonlinear simulations, and experiments on the chlorite-tetrathionate reaction.
\end{abstract}
\maketitle

Traveling fronts are commonly encountered in nature and applications, including combustion \cite{cla16}, nanoparticle self-assembly \cite{Bohner2015}, frontal polymerization \cite{poj,suslick2023,gao23}, emergence of homochirality \cite{Gillet2024} or subsurface and contaminant hydrology \cite{Rolle2019}. In biological systems, moving fronts control dynamics \cite{Murray2002} such as neuronal impulses \cite{Richardson2005}, plankton growth \cite{Semplice2013}, and disease spreading \cite{Ahmed2021}, to name a few. Fronts can develop when a local autocatalytic process is coupled to a diffusive transport and are characterized by a reactive zone moving with constant width and speed $c$ in space and time \cite{poj, Horvath1993, Scott1994book}. Transverse cellular modulations, resulting from diffusive or hydrodynamic instabilities, can deform such fronts. Convective deformations are typically caused by changes in composition and temperature across the front that trigger buoyancy-driven motions \cite{dew20} or Marangoni flows \cite{Tiani2018}, as seen in frontal polymerization \cite{gao23}, redox \cite{boe00, dew01}, and combustion fronts \cite{Radisson2021} for instance. Diffusive instabilities, by contrast, emerge when the reactant and autocatalyst have unequal diffusivities, and have been reported in redox \cite{Scott1992, Horvath1993} and combustion fronts \cite{siv77}.

Recently, it has been shown that radial injection can freeze autocatalytic fronts, opening perspectives in controlling pattern formation around such frozen fronts. Specifically, for an autocatalytic reaction of the type $ 2X + Y \rightarrow 3X $ where $X$ is the autocatalyst and $Y$ the reactant, stationary fronts can be obtained when the reactant $Y$ is injected radially into a medium initially filled with the autocatalyst $X$ \cite{Luka_2024}. The stationary radius $r_s$ is reached when the outward advection speed equals the inward reaction-diffusion (RD) speed. The fixed position $r_s$, the related curvature $1/r_s$, and the advective fluxes are all controlled by the flow rate, which paves the way to an advective control of instabilities that can develop around the front. 

Our objective is to test this new pattern-forming strategy both theoretically and experimentally by studying how radial advection modifies the diffusive instability of autocatalytic fronts.
Diffusive instabilities of autocatalytic fronts occur when the reactant $Y$ ahead of the front diffuses sufficiently faster than the autocatalyst $X$ invading it. To understand the mechanism of this instability, recall that, in convex perturbed regions ahead of the planar front, diffusion perpendicular to the front concentrates $Y$ while it dilutes $X$ (Fig.1a). In concave perturbations lagging behind the planar position, $X$ is concentrated and $Y$ is diluted. The diffusion of $X$ smooths out perturbations and stabilizes the front by spreading the autocatalyst from regions of high to low concentration, while the faster diffusion of $Y$ supplies more reactant to convex regions, locally accelerating the reaction and amplifying perturbations. If $D_Y>D_X$, where $D_X$ and $D_Y$ are the diffusion coefficients of $X$ and $Y$, respectively, the destabilizing diffusion of Y overcomes the stabilizing diffusion of X and perturbations amplify, leading to cellular patterns \cite{Scott1992, Horvath1993}. The diffusivity ratio $\delta=D_{Y}/D_{X}$ must exceed a critical value $\delta_{cr}$ to trigger cellular pattern formation around a traveling front \cite{Scott1992, Horvath1993, Zhang1994, Horvath1995, Milton1996, Merkin_2005}. 

How is this classical picture modified in radial advection? As Fig.1b shows, in a circular front expanding radially, the inner species is diluting when diffusing outward, while the inward diffusion of the outer species is concentrating it in the reference circle. The curvature inherent to radial geometries should thus naturally modify the $\delta_{cr}$ values of rectilinear systems. For a frozen front locked by advection at a position $r_s$, this geometrical effect should, in addition, be modulated by advection.

In this context, we study here theoretically the influence of geometric and advective effects on the diffusive instability of fronts, showing that new radially locked and angularly shifting sun-ray-like patterns can be obtained in radial reaction-diffusion-advection (RDA) systems.  Linear stability analysis (LSA) and direct nonlinear simulations (DNS) of an RDA model show that the critical diffusivity ratio $\delta_c$ of radial systems, as well as the amplitude and wavelength of the transverse modulations, are controlled by the flow rate $Q$. Experimental observations confirm the existence and properties of these star-shaped structures.

We consider a 2D circular domain defined in a coordinate system $(\hat{r}, \hat{\theta})$ with center at $\hat{r} = 0$ and radius $R$ and initially filled by the autocatalyst $X$.
The reactant $Y$ is injected radially from the center at a constant flow rate $\hat{Q}$  into the sea of  $X$. Both $X$  and $Y$ have the same initial concentration $\hat{Y_0}$ and undergo an autocatalytic reaction $2X + Y \rightarrow 3X$. The non-dimensional RDA equations describing the dynamics are:

\begin{subequations}\label{eq1}
\begin{gather}
    \partial_t x + \frac{Q}{r} \partial_r x = \left( \partial_{rr} + \frac{1}{r} \partial_r + \frac{1}{r^2} \partial_{\theta\theta} \right) x + x^2 y,\\
    \partial_t y + \frac{Q}{r} \partial_r y = \delta \left( \partial_{rr} + \frac{1}{r} \partial_r + \frac{1}{r^2} \partial_{\theta\theta} \right) y - x^2 y, 
\end{gather}
\end{subequations}
where $x$ and $ y$ are the concentrations of $ X$ and $ Y$, scaled by $ \hat{Y}_0$. The characteristic time and length scales are $ \hat{\tau} = 1 / (\hat{k} \hat{Y}_0^2)$ and $ \hat{l} = \sqrt{D_X \hat{\tau}}$, respectively, where $ \hat{k}$ is the kinetic constant. Hats refer to dimensional variables. The dimensionless advection term $Q/r$, with $Q = \hat{Q}/(2 \pi \hat{h} D_X)$, derives from the radially varying velocity field $\vec{u} = \left( u(\hat{r}) = \hat{Q}/{2 \pi \hat{h} \hat{r}}, 0 \right)$, where $ \hat{h}$ is the reactor height. The flow satisfies the incompressibility condition $ \nabla \cdot \vec{u} = 0$. Eqs. (\ref{eq1}) are solved numerically with initial conditions $ (x,y)(r, \theta, 0) = (0,1)$ at $r = r_0$ and $(1,0)$ for $r_0 < r \leq R$.
The boundary conditions are: \(x(r_0, \theta, t) = y(R, \theta, t) = 0\), and \(x(R, \theta, t) = y(r_0, \theta, t) = 1\). To avoid the singularity at the center, the numerical domain is defined over $r \in [r_0, R]$ with $r_0 = 0.5$ and $R = 700$. In presence of radial advection, the RDA eqs.(1) admit a stationary front at a position $r_s=Q/c$ where the outward velocity $u=Q/r$ balances the inward RD front speed $c$ \cite{Luka_2024}. 

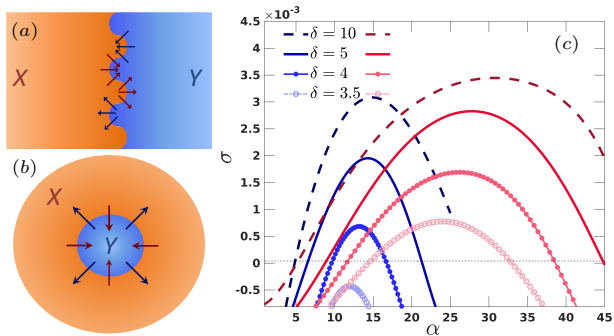
\begin{figure}[h!]
\includegraphics[width=0.32 \columnwidth]{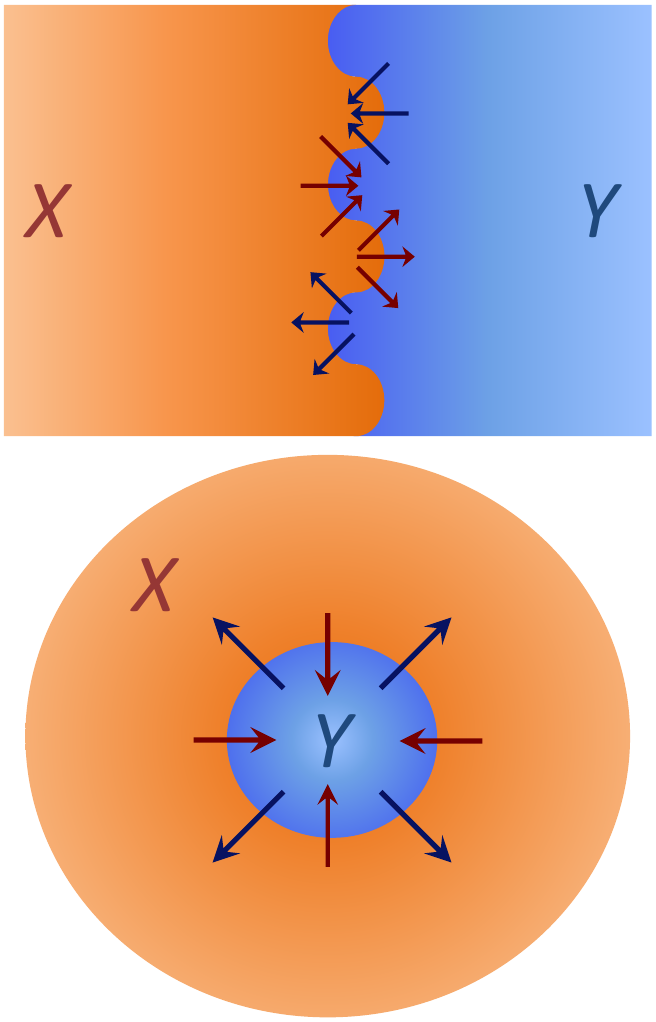}
\includegraphics[width=0.6 \columnwidth]{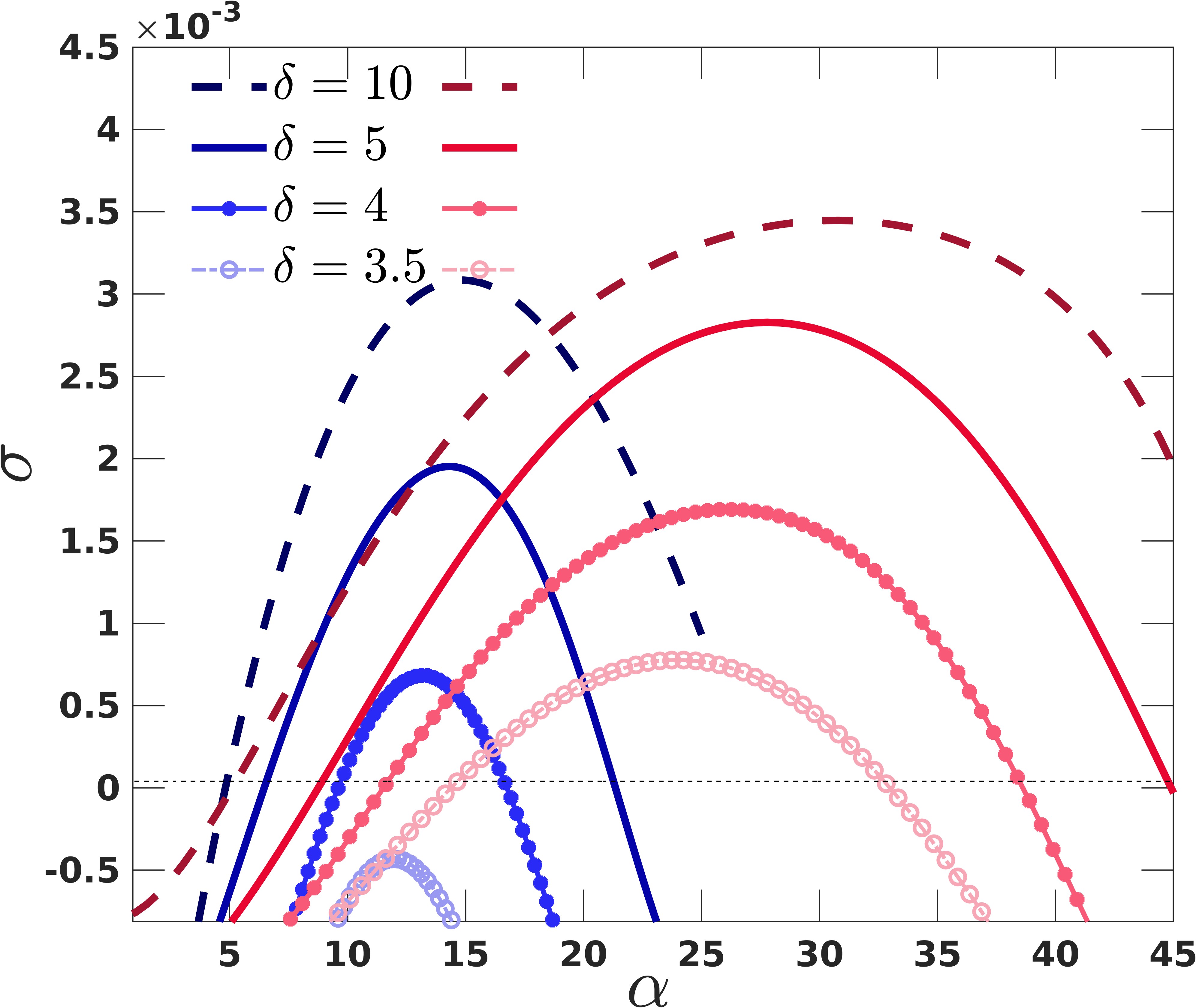}
    \begin{picture}(0,0)
    \put(-225,115){\makebox(0,0)[]{\scriptsize $(a)$}}
    \put(-225,63){\makebox(0,0)[]{\scriptsize $(b)$}}
    \put(-20,110){\makebox(0,0)[]{\scriptsize $(c)$}}
   \end{picture}
\vspace{-.4 cm}\caption{Direction of diffusion in (a) rectilinear and (b) radial geometries. (c) Growth rate $\sigma$ versus wave number $\alpha$ for $Q=80$ (blue) and $Q=160$ (red) at various values of $\delta$.
\label{fig_LSA} 
}
\end{figure}

To analyze diffusive instabilities of such locked fronts, we first perform a temporal LSA \cite{llamoca2022} of eqs.~\eqref{eq1} to understand the influence of $Q$ and of the related curvature $1/r_s$ on the instability.
We compute dispersion curves giving the growth rate $\sigma$ of angular $cos(\alpha \theta)$ perturbations around the expanding circular front as a function of their angular wavenumber $\alpha$.  See the supplemental material (SM) for details \cite{supplemental}. Fig.1c shows that, depending on the flow rate $Q$ and diffusivity ratio $\delta$, perturbations either grow ($\sigma > 0$) or decay  ($\sigma < 0$). 
The maximum growth rate $\sigma_{max}$ and the most unstable wavenumber $\alpha_{max}$ increase when the diffusivity ratio $\delta$ is increased, as known for both rectilinear \cite{Scott1992, Horvath1993} and radial RD fronts \cite{Milton1996}. 
Increasing $Q$ raises the value of $\alpha_{max}$ \textit{i.e.} destabilizes smaller angular perturbations. At the same time, it shifts the front outward, to a larger $r_s$. As a result, the wavelength $\lambda_{max} = 2 \pi r_s/\alpha_{max}$ increases only slightly.

For example, with $\delta = 10$, increasing $Q$ from 80 to 160 nearly doubles $\alpha_{max}$ from approximately 15 to 30 (Fig.~\ref{fig_LSA})c), while the front shifts outward from $r_s \approx 230$ to $r_s \approx 460$ (see Fig.S3 of \cite{supplemental}). Consequently, the most unstable wavelength changes only slightly, from around 95 to 97.
Conversely, at a fixed $Q$, increasing  $\delta$ leads to a modest rise in $\alpha_{max}$ but decreases the RD speed $c$, which shifts the front to a larger radial position $r_s$, resulting in a larger wavelength. For instance, with $Q = 160$, increasing $\delta$ from 3.5 to 10 increases $\alpha_{max}$ from 25 to 30 but shifts $r_s$ from $\sim$ 305 to 460 (Fig. S3  \cite{supplemental}), increasing $\lambda_{max}$ from around 77 to 97. This highlights how increasing $\delta$ promotes more widely spaced instability patterns.
Overall, increasing $Q$ and $\delta$ both enhance instability. Increasing $\delta$ primarily affects the wavelength, while increasing $Q$ impacts rather the critical diffusivity ratio.

\begin{figure}[h!]
\includegraphics[scale=0.32]{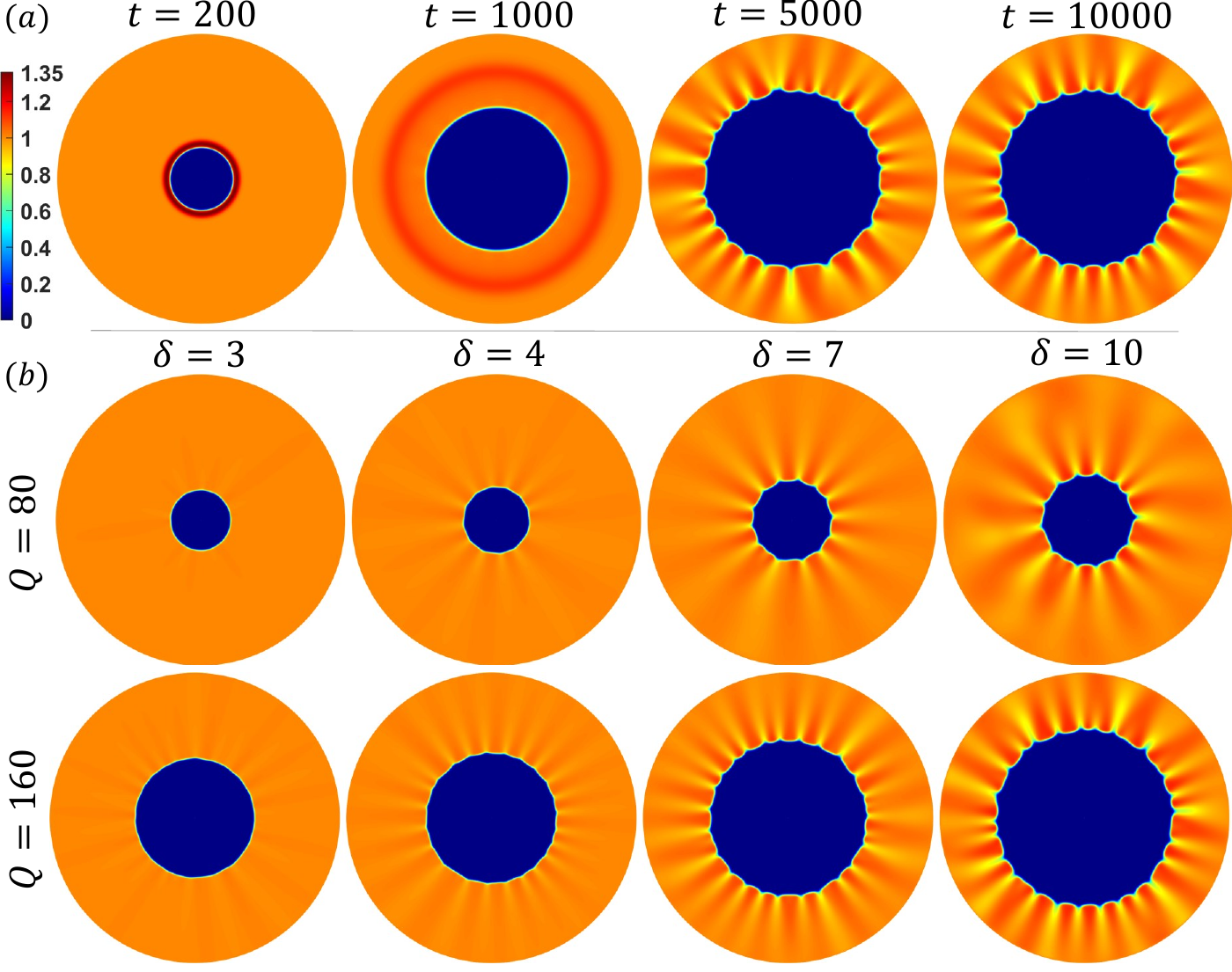}
\vspace{-.3 cm}\caption{\label{fig2} (a) Spatio-temporal evolution of the radial displacement of the autocatalyst $X$ (in orange) by the reactant $Y$ (in blue) for $Q=160$ and $\delta=10$. (b) Distribution of $X$ at $t=10000$ for  $Q=80$ or 160 and variable  $\delta$.}
\end{figure}
\begin{figure*}\includegraphics[width=\textwidth]{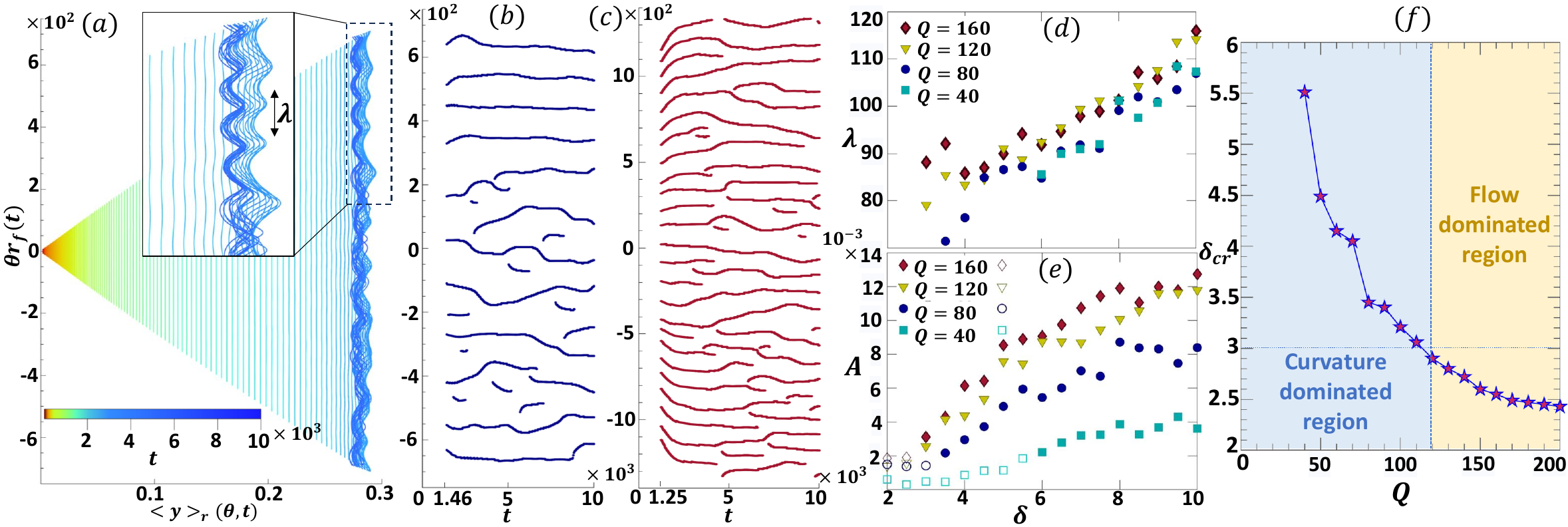}
\vspace{-.5 cm}\caption{\label{fig3} (a) Temporal evolution of the $r$-averaged reactant concentration, $\langle y \rangle_r(\theta,t)$ for $Q=80$ and $\delta=10$. Inset: Zoom showing the unstable interface receding towards the center and becoming radially locked. Space-time maps of sun-ray source points for (b) $Q=80$ and (c) $Q=160$. (d) Mean wavelength $\lambda$ and (e) amplitude $A$ of the cellular modulations for variable $Q$ and $\delta$. Empty and filled markers represent stable and unstable ($Q,\delta$) pairs, respectively. (f) Critical diffusivity ratio $\delta_{cr}$ versus flow rate $Q$. The line at $\delta_{cr}$=3 is the value of the rectilinear case \cite{Horvath1993}.}
\end{figure*} 

Next, we perform nonlinear simulations of the RDA model to validate the LSA predictions and analyze nonlinear dynamics. Details of the numerical method and validation are provided in the SM \cite{supplemental}. Once locked at $r_S$, the sun-ray sources feature a complex nonlinear dynamics involving lateral movements, merging, or birth of new rays (see Movie 1).
Fig.\ref{fig2}(a) shows the temporal evolution of the radial displacement of the autocatalyst $X$ by the reactant $Y$. Since $\delta > 1$, the faster diffusion of $Y$  allows   $X$ to locally exceed unity, forming a $\theta$-symmetric circular second diffusion-advection front \cite{Luka_2024} emerging from the reactive stationary front to adapt the initial outer concentration of X to its stoichiometry ratio with Y  (see Fig.\ref{fig2}(a) at $t = 200, 1000$ and Fig. S8(a) in SM \cite{supplemental}). A diffusive instability may deform the reaction front behind it into sun-ray-like patterns either before or after this radial locking at the position $r_s$  (See Fig.\ref{fig2}(a) at $t = 5000, 10000$ and Fig.S8(a,b) \cite{supplemental}). Fig.\ref{fig2}(b) illustrates the impact of varying $Q$ and $\delta$ on the sun-ray patterns. At a fixed flow rate, increasing the diffusivity ratio amplifies the diffusive instability, with sun-rays appearing earlier and their asymptotic amplitude being larger. At a fixed value of $\delta$, the system becomes more unstable, with larger amplitudes of the cellular modulations, as $Q$ increases.  

Fig.\ref{fig3}(a) shows the time evolution of the $r$-averaged $Y$ concentration, $\langle y \rangle_r(\theta,t) = \frac{1}{R-r_0} \int_{r_0}^{R} y(r,\theta,t) \, dr$, as a function of $\theta r_f(t)$, the interfacial perimeter. Here, $r_f(t)$ is the reaction front position defined as the first $r$-moment of the reaction rate $\mathcal{R}=x^2y$, \textit{i.e.} 
\(r_f(t) = \frac{\int_{r_0}^{R} r \langle \mathcal{R} \rangle_{\theta}(r, t) dr}{\int_{r_0}^{R} \langle \mathcal{R} \rangle_{\theta}(r, t) dr},\)
where  $\langle \mathcal{R} \rangle_\theta(r,t) = \frac{1}{2 \pi} \int_{0}^{2\pi} \mathcal{R}(r,\theta,t) \, d\theta$.
Fig.3a shows that the initially circular interface grows uniformly, increasing the front perimeter. Eventually, the front destabilizes transversely, while stopping its growth.
Space-time maps of the dynamics can next be constructed by stacking in time the position of the local minima of $\langle y \rangle_r(\theta,t)$(Fig.\ref{fig3}(b,c)). These maps display the angular movement of the sun-ray sources on the interfacial perimeter as well as nonlinear dynamics such as the birth or merging of rays. We mark the onset time $t_{on} = \min \{t : A(t) > 2 \cdot 10^{-3} \}$ of instability, where 
$A(t) = \underset{\theta}{\max} \left| \langle y \rangle_r(\theta, t) - \langle \langle y \rangle_r(\theta, t) \rangle_\theta \right|$ 
is the maximum amplitude of the transverse modulations seen in Fig.\ref{fig3}(a). The instability occurs earlier at larger flow rates (e.g. the dimensionless onset time is $\approx 1250$ for $Q=160$ while $\approx 1467$ for $Q=80$), but a longer delay is observed between onset and radial locking. 
 This can be seen on the space-time maps of Fig.\ref{fig3}(b,c) starting at onset of the instability: the minima of the cellular modulations still expand after onset for $Q=160$ while the unstable interface shifts slightly inwards and the radial expansion of the unstable front until locking is less pronounced for $Q=80$ (see inset of Fig.\ref{fig3}(a) and Fig.S9(a) in SM \cite{supplemental}).

We next calculate the wavelength $\lambda(t) = \theta r_f(t)/{n(t)}$ of the transverse modulations as the ratio between the interfacial perimeter $\theta r_f(t)$ at the front position divided by $n(t)$, the number of local minima. Because of the nonlinear interactions between rays, $\lambda(t)$ fluctuates around a mean $\lambda$ (see Fig.S9 (b) of SM \cite{supplemental}). Fig.\ref{fig3}(d) shows that this mean instability wavelength increases with  $\delta$, while it does not vary much with $Q$, in agreement with the LSA predictions.
Similarly, the amplitude $A(t)$ of the front modulations first rises after onset of the instability and then plateaus, fluctuating around a mean value $A$ (see Fig.S9(c) of SM \cite{supplemental}). Fig.\ref{fig3}(e) shows that this mean amplitude $A$ increases, \textit{i.e.}, the system is more unstable when $\delta$ and $Q$ increase, which is again consistent with the LSA predictions.

For each value of $Q$, we define the critical diffusivity ratio $\delta_{cr}$ as the value of $\delta$ above which the mean amplitude $A >  2 \cdot 10^{-3}$. 
Fig.\ref{fig3}(f) compares the LSA and DNS  values of $\delta_{cr}(Q)$ to the DNS value $\delta_c^{\ \text{Rect}}=3.0$ obtained numerically for rectilinear RD fronts. 
Both DNS and LSA demonstrate that radial flow conditions can modify the critical threshold of instability. 
\begin{figure*}
\includegraphics[trim=0 30 0 0,clip,width=\textwidth]
{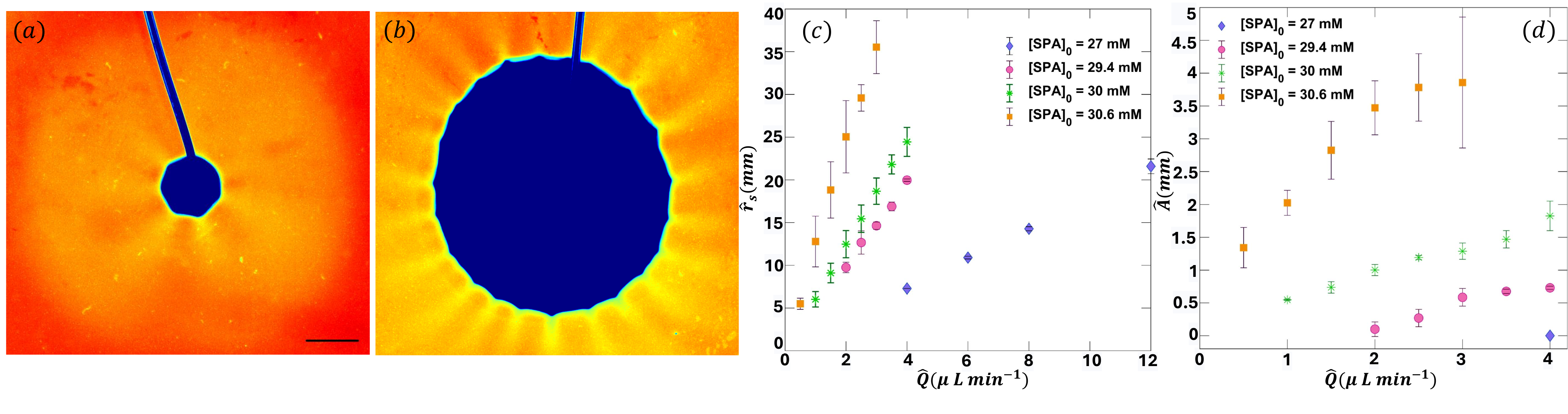}
\vspace{-.5 cm}\caption{\label{fig4} Cellular front of the CT reaction at [SPA]$_0 =$ 30 mM (Scale bar = 10 mm) for (a) $\hat{Q} = 1$  $\mu$ L min$^{-1}$, and (b) $\hat{Q} = 4$ $\mu$ L min$^{-1}$. (c) Stationary front position $\hat{r}_s$ (mm),  and (d)  amplitude $\hat{A}$ (mm) of the cellular modulation versus flow rate $\hat{Q}$ for different [SPA]$_0$.}
\end{figure*}

If we focus on the DNS values, two regimes can be identified. For lower $Q$, the critical threshold is larger than the rectilinear value ($\delta_{cr}^{Rect} \approx 3$). This can be understood by the fact that, at a smaller flow rate, the stationary radius $r_s$ is smaller and hence curvature larger. As seen on Fig.1b, this curvature effect dilutes Y out of the inner circle, while it concentrates X inside it. The destabilizing role of Y is weakened compared to the enhanced stabilizing influence of X and the net critical ratio $\delta_{cr}$ needed to destabilize the front is larger. The frozen fronts with a small curvature are thus more stable with regard to the diffusive instability than the traveling rectilinear ones, in the so-called curvature-dominated region of the $(\delta_{cr},Q)$ parameter space (Fig.3f). At larger $Q$, the curvature effect weakens as the stationary radius is larger. There, it is useful to note that the total flux $\vec{F}=\vec{F}_d+\vec{F}_{adv}$ of a given species at any location is the sum of its diffusive $\vec{F}_d$ and advective $\vec{F}_{adv}$ components. At the stationary front position, for species Y, these two components align in the same direction, while for species X, the diffusive flux is in the direction opposite to the advective one. We have thus naturally that the amplitude of the total Y flux is larger than the X one, and this difference increases with $Q$, favoring a decrease of $\delta_{cr}$ in the flow-dominated region of the $(\delta_{cr},Q)$ parameter space (Fig.3f). Beyond a certain $Q$, the diffusive component becomes negligible and $\delta_{cr}$ saturates. For large $Q$, the DNS $\delta_{cr}$ asymptotically decreases towards $\delta_{cr} = 2.4$, which is below $\delta_{cr}^{Rect}$.

To test theoretical predictions, experiments are performed with the chlorite-tetrathionate (CT) reaction \cite{Horvath1998, RICA2010831} where the autocatalyst $X$ are protons. The reactive solution $Y$ consists of 5 mM potassium tetrathionate (K$_2$S$_4$O$_6$), 20 mM of sodium chlorite (NaClO$_2$), 1 mM sodium hydroxide (NaOH), a pH indicator, and variable initial concentrations of sodium polyacrylate ($27$ mM $\leq$ [SPA]$_0$ $\leq 30.6$ mM). The solution $X$ is obtained by adding a droplet of acid to the solution $Y$, initiating the reaction. Carboxyl groups of SPA reversibly bind the autocatalyst, effectively decreasing its diffusive and reactive timescales \cite{RICA2010831, Jakab2001}. [SPA]$_0$ plays in experiments a role equivalent to that of $\delta$ in the simulations.

The solution $X$ initially fills a horizontal Hele-Shaw cell, two glass plates separated by a thin gap ($\hat{h}=0.12$ mm). The solution $Y$ is injected radially from a hole in the center of the cell at a constant flow rate $\hat{Q}$ using a syringe pump. The evolution of the front is recorded using a digital camera (see Fig. S6 in SM \cite{supplemental} for a schematic of the experimental set-up). The experimental data are processed as described in SM \cite{supplemental}.

For given values of $\hat{Q}$ and [SPA]$_0$, above a critical value of [SPA]$_0$, the front develops cellular modulations, providing shining star-like structures at the locked radius.  This pattern has a given number of cusps which drift laterally, can merge, while new ones sometimes appear locally (see Movie 2). Increasing $\hat{Q}$ at constant [SPA]$_0$ results in a front advected farther from the center (see Fig.4c) with decreased curvature and thus a total flux of Y larger than that of X, allowing for a larger amplitude of cellular modulation. This behavior is demonstrated in Fig.4a and 4b, where the modulation amplitude increases from 0.5 mm to 1.7 mm when changing $\hat{Q}$ from 1 to 4 $\mu$L min$^{-1}$. Fig.4d shows that increasing $\hat{Q}$ destabilizes the system, \textit{i.e.}, leads to larger amplitudes of modulations, for each value of [SPA]$_0$ scanned. 

Increasing [SPA]$_0$ results in more bounded protons, hence a slower RD velocity $\hat{c}$ of the front, and a larger stationary front position $\hat{r}_s$ for the same $\hat{Q}$ (Fig.4c).  It also favors the diffusive instability, inducing larger cellular modulations of the front (Fig.4d). Unlike in simulations and LSA, the wavelength of the pattern is roughly constant ($8 \pm 2$) mm \textit{i.e.} does not vary with the flow rate or [SPA]$_0$. The fact that the wavelength is independent of [SPA]$_0$ is consistent with previous findings \cite{RICA2010831}.

To summarize, we have experimentally and theoretically demonstrated the existence of new sun-ray-like patterns in self-organized autocatalytic RDA systems, resulting from a diffusive instability of advectively frozen fronts. Their properties can be tuned by varying the flow rate. At low flow rates, the curvature-dominated regime has a larger critical diffusivity ratio than in rectilinear systems while at larger $Q$, the frozen radial fronts are more unstable than their rectilinear counterparts because advection becomes dominant over diffusion. These results show how radial advection of autocatalytic fronts can be used for controlling pattern formation by freezing a pattern at a given radius which controls the relative weight of curvature and transport fluxes on the dynamics. This paves the way to discover new self-organized patterns around frozen fronts, in particular due to buoyancy or viscosity effects \cite{dew20}. 
 
S.M. and L.N. contributed equally to this work. Financial support from the Actions de Recherches Concert\'ees CREDI, from ULB and from Prodex is acknowledged. 
\bibliographystyle{apsrev4-1}
\bibliography{Main}

\end{document}


\pagebreak
\widetext
\begin{center}
\underline{\textbf{\large Supplemental Material}}

\vspace{0.2cm}

\textbf{\large {Radially Locked Sun-Ray Patterns in an Autocatalytic Reaction-Diffusion-Advection System}}\\
\vspace{0.2cm}
S. N. Maharana, L. Negrojević, A. Comolli and A. De Wit\\
\vspace{0.1cm}
Nonlinear Physical Chemistry Unit, Université libre de Bruxelles, 1050 Brussels, Belgium
\end{center}

\setcounter{equation}{0}
\setcounter{figure}{0}
\setcounter{table}{0}
\setcounter{page}{1}
\makeatletter
\renewcommand{\theequation}{S\arabic{equation}}
\renewcommand{\thefigure}{S\arabic{figure}}
\renewcommand{\bibnumfmt}[1]{[#1]}
\renewcommand{\citenumfont}[1]{#1}
\maketitle

\section{The linear problem, growth rate computation and numerical method Validations}

We perform a temporal linear stability analysis (LSA) to analyze instability dynamics in the linear regime \cite{llamoca2022}. Assuming that the concentrations \( x \) and \( y \) are \( \theta \)-periodic, they can be approximated by the leading terms of their Fourier series expansions as: 
\begin{subequations}\label{eq2}
\begin{gather}
     x(r, \theta, t) \approx x_0(r, t) + x_1(r, t) \cos(\alpha \theta),\\
     y(r, \theta, t) \approx y_0(r, t) + y_1(r, t) \cos(\alpha \theta),
\end{gather}
\end{subequations}
where \( x_0(r, t) \) and \( y_0(r, t) \) profiles represent the axisymmetric base concentration profiles, and \( x_1(r, t) \) and \( y_1(r, t) \) are the amplitudes of transverse modulations along \( \theta \). Here, \( \alpha \) is the angular disturbance wavenumber associated with the Fourier mode. The corresponding physical wavelength is given by $\lambda=\frac{2 \pi r}{\alpha}$. Substituting Eqs.~\eqref{eq2} into the RDA Eqs. (1) of the article, and using a first-order Taylor expansion of the rate law \( x^2 y \) around \( x_0(r, t) \) and \( y_0(r, t) \), yields:
\begin{subequations}\label{eq3}
\begin{gather}
  \partial_t x_0 =  \left( \partial_{rr} x_0 + \frac{1-Q}{r} \partial_r x_0 \right) + x_0^2 y_0,\\
  \partial_t y_0 = \delta\!\left( \partial_{rr} y_0 + \frac{1-Q}{r} \partial_r y_0 \right) - x_0^2 y_0,\\
  \partial_t x_1 =  \left( \partial_{rr} x_1 + \frac{1-Q}{r} \partial_r x_1 - \frac{\alpha^2}{r^2} \right)\!x_1 + \mathcal{R}_{lin},\\
  \partial_t y_1 = \delta\!\left( \partial_{rr} y_1 + \frac{1-Q}{r} \partial_r y_1 - \frac{\alpha^2}{r^2} \right)\!y_1 - \mathcal{R}_{lin},
\end{gather}
\end{subequations}
where \( \mathcal{R}_{lin} = x_0^2 y_1 + 2 y_0 x_0 x_1 \). The initial conditions for the linear system (\ref{eq3}) are given by: \( y_0(r, 0) = 0 \) for \( r_0 \leq r \leq r_I \), and \( y_0(r, 0) = 1 \) for \( r_I \leq r \leq R \); similarly, \( x_0(r, 0) = 1 \) for \( r_0 \leq r \leq r_I \), and \( x_0(r, 0) = 0 \) for \( r_I \leq r \leq R \). The initial radial location \( r_I \) is set to \( r_I = 200 \) to avoid boundary effects. For the perturbations at \( r = r_I \), we set \( y_1(r, 0) = +0.001 \) and \( x_1(r, 0) = -0.001 \) \cite{llamoca2022}.
Dirichlet boundary conditions consistent with the non-linear problem are applied, i.e., \( y_0 = 1 \) at \( r = r_0 \), \( y_0 = 0 \) at \( r = R \), \( x_0 = 0 \) at \( r = 0 \), \( x_0 = 1 \) at \( r = R \).
The linear problem (\ref{eq3}) is numerically solved for \( t \in [0, 10000] \) using the Finite Element Method (FEM) via LiveLink\texttrademark\ for \textsc{Matlab}\textsuperscript{\textregistered} and \textsc{COMSOL Multiphysics}\textsuperscript{\textregistered} over the interval \( r \in [0.5, 700] \). The Coefficients of PDE module is selected from the equation-based modeling component, with an element size of 0.025. The time-dependent solver is set as a direct solver using the PARDISO linear solver. For time stepping, the Backward Difference Formula (BDF) is applied with an initial step size \( \Delta t = 0.001 \). 
\begin{figure}[h!]
\includegraphics[width=0.6\columnwidth]{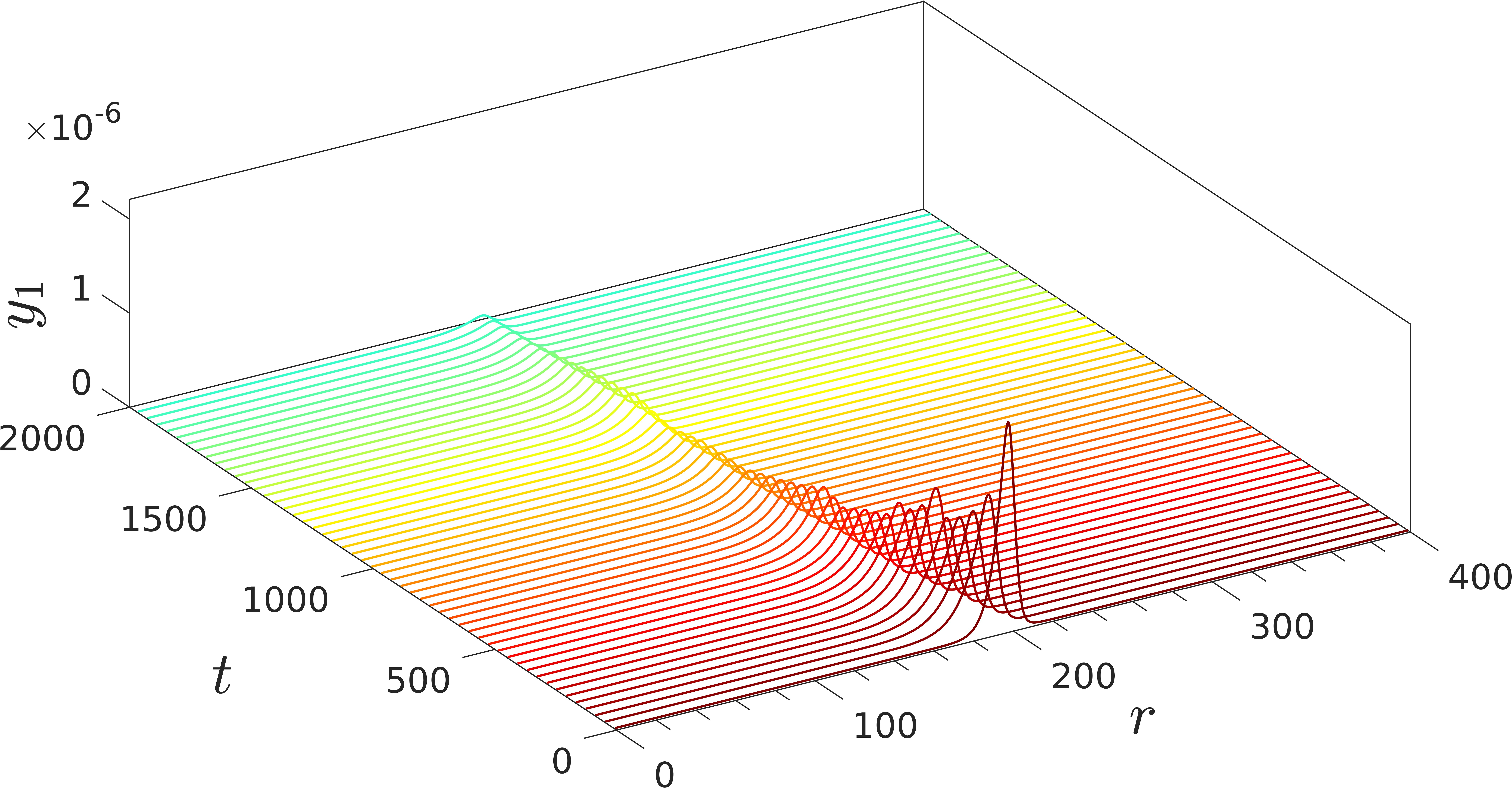}
\includegraphics[width=0.6\columnwidth]{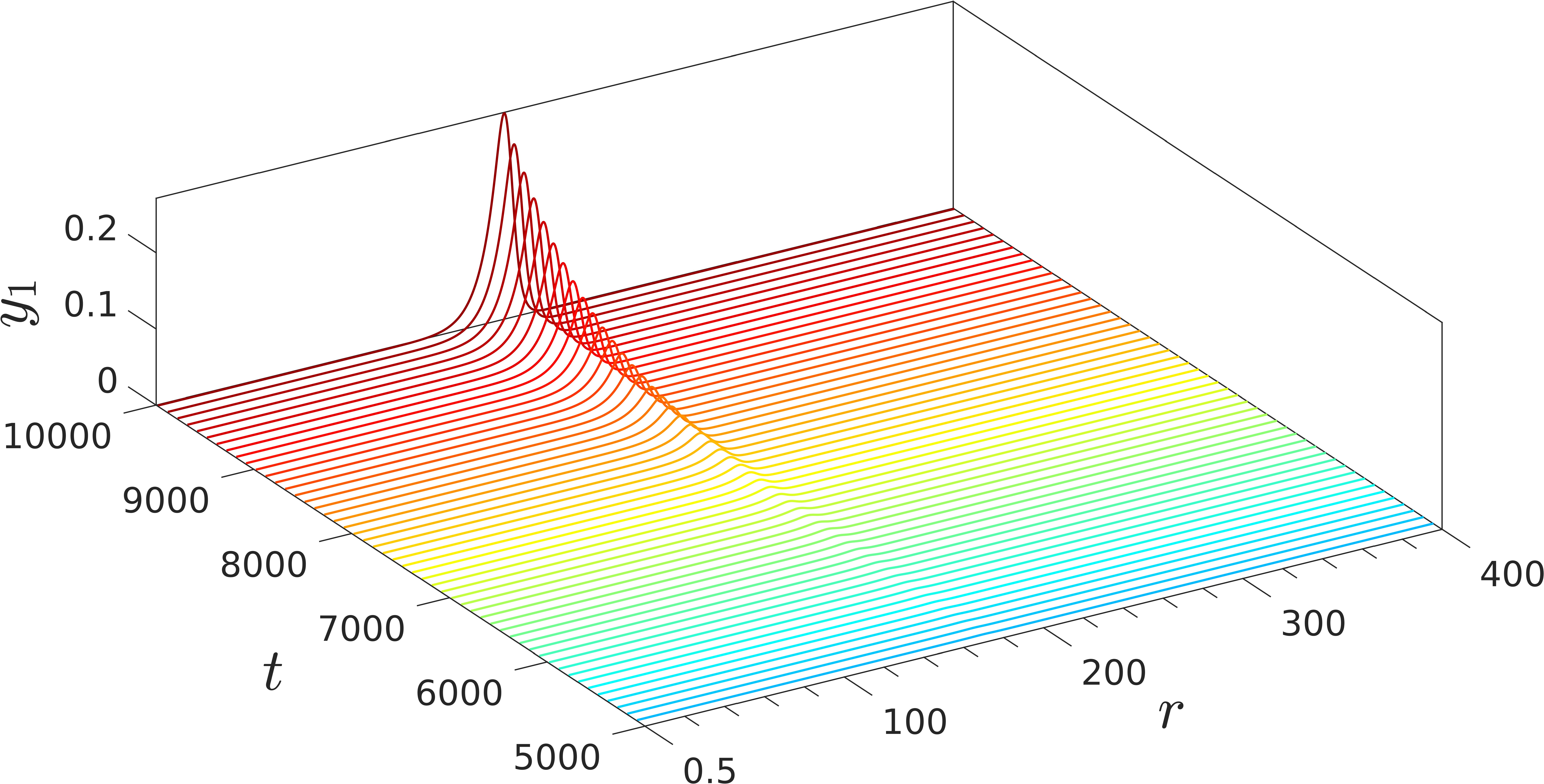}
\vspace{-.5 cm}\caption{\label{fig_LSA}  Spatio-temporal profiles of perturbation $y_1$ showing that the mode with (a) $\alpha=2$ decays over time, while the mode with (b) $\alpha=10$ grows over time for $Q=80$ and $\delta=5$.
}
\end{figure}

Fig.~\ref{fig_LSA}(a) shows that for an angular disturbance wavenumber \( \alpha=2 \), the perturbation \( y_1 \) decays over time with \( Q=80 \) and \( \delta=5 \). However, for \( \alpha=10 \), \( y_1 \) grows with the pertubation, time (Fig.~\ref{fig_LSA}(b)). The growth or decay of perturbations, seen in Fig.~\ref{fig_LSA}(a, b), occurs at a fixed position \( r \) over time, showing that, in the base system, the competition between the consumption of reactant \( Y \)  by the autocatalyst \( X \) from \( r > r_I \) toward the center and outward advection of \( Y \) from the center radially locks the front. The perturbations thus decay or grow around this radially-locked front.
Growth rates \( \sigma \) for each \( \alpha \) are determined by least squares, using the maxima of \( y_1 \), denoted \( y_{\max_i} \), and discrete times \( t_i \), where \( i=1, \ldots, m \) and \( m \) is the number of maxima \cite{llamoca2022}:
\[
\sigma = \frac{m \sum t_i \log y_{\max_i} - \sum t_i \sum \log y_{\max_i}}{m \sum t_i^2 - \left( \sum t_i \right)^2}.
\]
This growth rate can also be calculated from \( x_1 \), as discussed in \cite{llamoca2022}. We have reproduced Figure 2 from \cite{llamoca2022}, which aligns perfectly with our computations.

\section{Supplemental results from linear stability analysis}
\begin{figure}[h!]
	\centering
	\includegraphics[scale=.07]{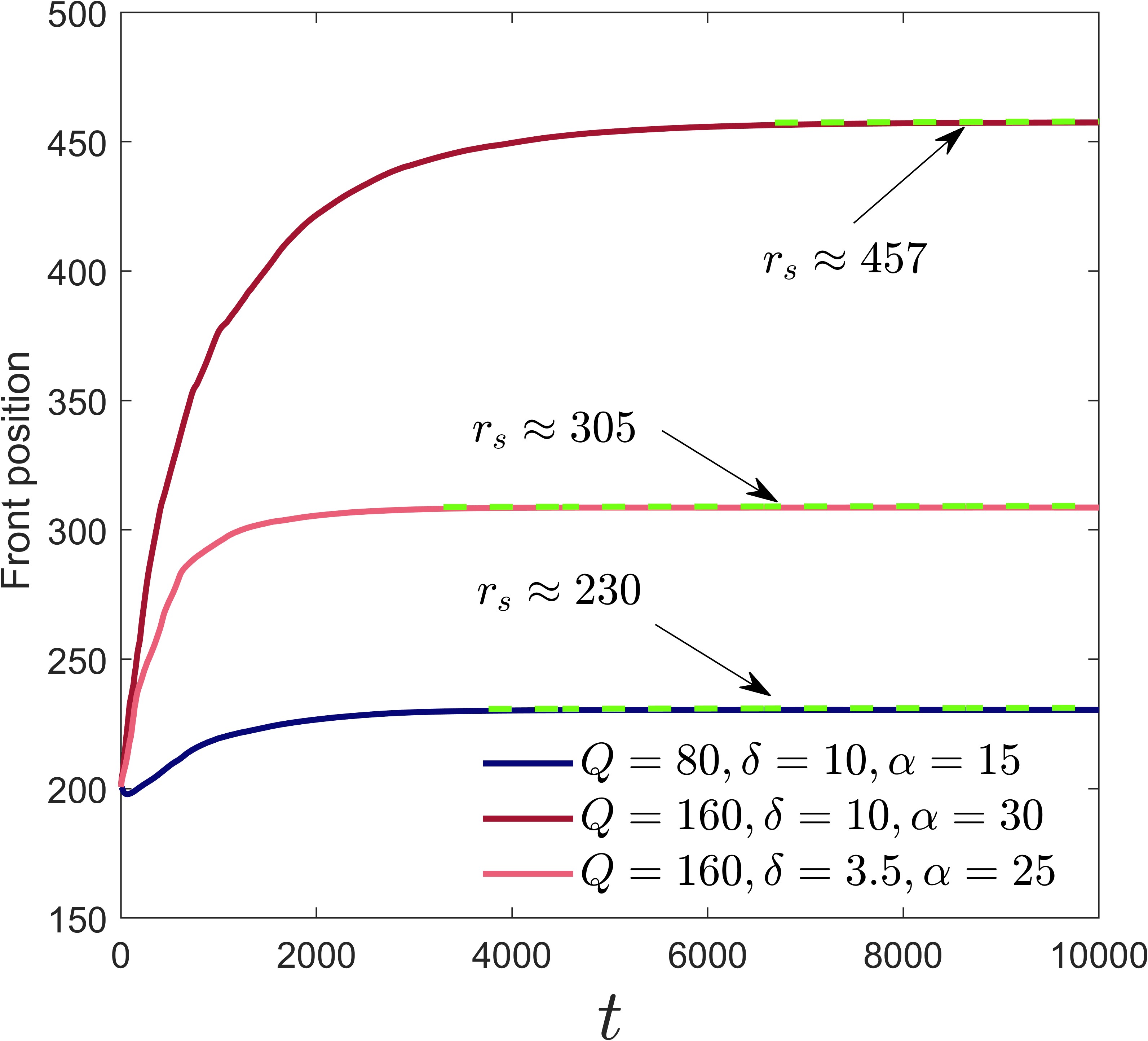}
	\caption{{\it The reaction front position evolution for different flow rate $Q$, diffusivity ratio $\delta$ and disturbance wavenumber $\alpha$. The green dashed lines indicate corresponding stationary front position $r_s$.}}
	\label{rf_LSA}
\end{figure}
The reaction front position in the base system, defined as \( r_f^{base} = \frac{\int_{r_0}^R r \, x_0^2 \, y_0 \, dr}{\int_{r_0}^R x_0^2 \, y_0 \, dr} \), \textit{i.e.} the first moment of the reaction rate $R=x^2y$, is plotted in Fig.S2 for different flow rates (\( Q \)), diffusivity ratios (\( \delta \)), and angular disturbance wavenumbers (\( \alpha \)). After an initial transient period, the front position becomes stationary. For \( Q = 80 \), \( \delta = 10 \), and \( \alpha = 15 \), the stationary front is at \( r_s \approx 230 \). For \( Q = 160 \), \( \delta = 10 \), and \( \alpha = 30 \), the stationary position increases to \( r_s \approx 305 \), while for \( Q = 160 \), \( \delta = 3.5 \), and \( \alpha = 25 \), it reaches \( r_s \approx 457 \).

\section{Numerical method for the nonlinear problem and validation}

\subsection{Numerical method}

The nonlinear problem is solved using the Finite Element Method (FEM) implemented in \textsc{COMSOL Multiphysics}\textsuperscript{\textregistered}, interfaced with \textsc{Matlab}\textsuperscript{\textregistered} via LiveLink\texttrademark, over a circular domain. The singularity at the centre is avoided by introducing an injection hole with a radius $r_{0}=0.5$. To neglect the boundary effects on the reaction dynamics, the computational domain is defined with an outer boundary at a radius of $R=700$, ensuring a significant distance from the reaction front for all used parameters in this study. The used FEM approach employs Delaunay tessellation to generate a regular triangulation, thereby preventing distortions in the pattern formations when interpolated \cite{Preparata1988,George1998,sharma2021}. The simulation uses COMSOL’s default adaptive time-stepping scheme, based on first- and second-order backward differentiation formulas (BDF) for temporal discretization. For the initial time step, a Backward Euler method with a fixed small step size of $\Delta t=10^{-3}$ is used to ensure result independence at the final time. Table \ref{tab1} presents the values or ranges of the parameters used in this study. Note that, for better visualization, we selectively zoom or crop the domain, adjusting the modified domain size to focus on specific aspects of interest.

\begin{table}[h!]
\centering
\begin{tabular}{cc}
\hline
\textbf{Parameters} & \textbf{Values} \\
\hline
$m$ & 0.1 \\
Number of Degrees of Freedom & 3270616\\
$r_0$ & 0.5 \\
$R$ & 700 \\
$Q$ & [10, 200] \\
$\delta$ & [1, 10] \\
\hline
\end{tabular}
\caption{Parameter values or ranges selected for this study.}
\label{tab1}
\end{table}
\begin{table}[h!]
\centering
\begin{tabular}{ccccccc}
\hline
\hline
$m$ & 0.05 & 0.1 & 0.2 & 0.4 & 0.8 & 1.6 \\
DoF &3272056&3270616&3243616&3220864&3204808&3201952\\
\hline
\end{tabular}
\caption{Various $h$ and corresponding Degrees of Freedoms (DoF) in the mesh independence test.}
\label{tab2}
\end{table}
 
\subsection{Mesh independence}
\begin{figure}[h!]
	\centering
	\hspace{0.45 in} (a) \hspace{3.05 in} (b)\\
	\includegraphics[scale=.6]{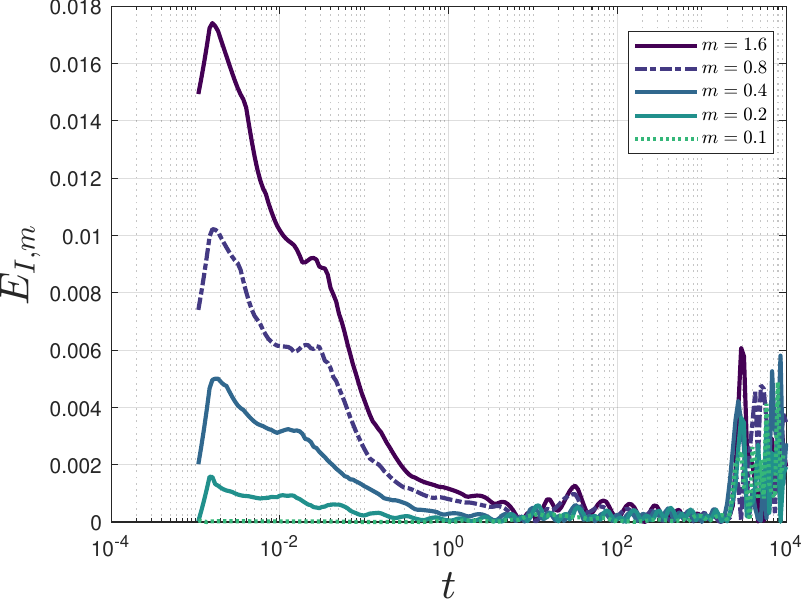}
	\includegraphics[scale=.6]{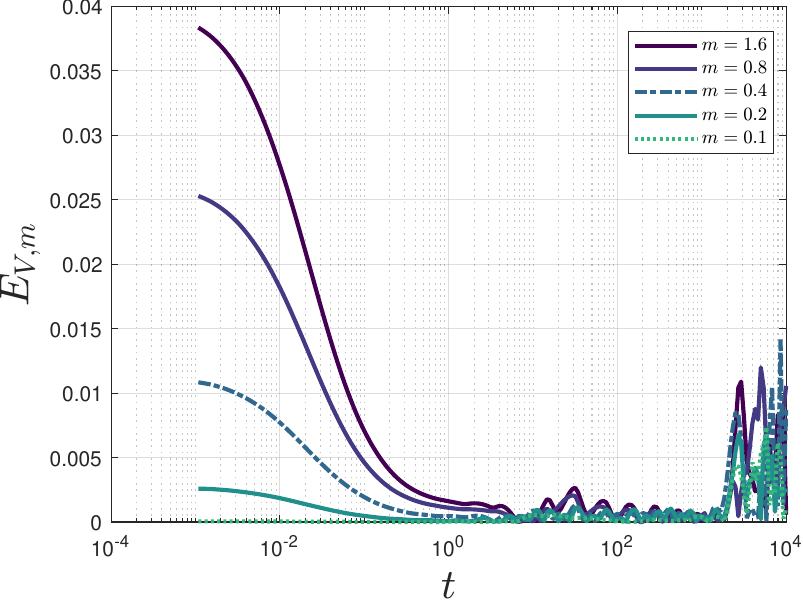}
	\caption{{\it  (a)  $E_{I,m}(t)$ and (b) $E_{V,m}(t)$ for different $m$ values. The rest of the parameters are fixed as $Q=80$, $\delta=10$.}}
	\label{mgridtest}
\end{figure}

The free triangular type finite element mesh (FEM) is selected from the fluid dynamics module for conducting simulations. The maximum element size is fixed at 5 units. To assess the mesh independence of the numerical solutions, the mesh is refined by varying the minimum element size, denoted as $m$, from the set $\{1.6, 0.8, 0.4, 0.2, 0.1, 0.05\}$ (corresponding degrees of freedom are given in Table \ref{tab2}) . For this purpose, we consider the solution obtained from the finest grid, $m = 0.05$, as our reference for the most accurate results and the computed quantities with this minimum element size is denoted by a subscript $e$.
Next, we calculate the relative error in the interfacial length, denoted as $E_{I,m}(t)$, and the total amount of the invading reactant, denoted as $E_{V,m}(t)$, with respect to other element sizes. These errors are defined as follows:
\begin{equation}
E_{I,m}(t) = \frac{\left| I_{m}(t) - I_{e}(t) \right|}{I_{e}(t)}, \quad E_{V,m}(t) = \frac{\left| V_{m}(t) - V_{e}(t) \right|}{V_{e}(t)}
\end{equation}
where the subscripts $m$ denote that the values are computed by varying $m$.
Here, the interfacial length, $I(t)$, is defined as:
\begin{equation}
I(t) = \int_{r_{0}}^{R} \int_0^{2\pi} \sqrt{\left(\frac{\partial y}{\partial r}\right)^2 + \frac{1}{r^2}\left(\frac{\partial y}{\partial \theta}\right)^2} r d\theta~ dr.
\end{equation}
Similarly, the total amount of the invading reactant fluid, $V(t)$, is given by:
\begin{equation}
V(t) = \int_{r_{0}}^{R} \int_0^{2\pi} y r d\theta~ dr.
\end{equation}


In Fig. \ref{mgridtest}(a) and Fig. \ref{mgridtest}(b), we present the plots of the relative errors $E_{I,m}(t)$ and $E_{V,m}(t)$ as functions of time for various minimum element sizes. As the minimum element size ($h$) decreases, both $E_{I,m}(t)$ and $E_{V,m}(t)$ decrease, indicating the convergence of the numerical solutions. Among the different sizes, the error is the smallest for $m=0.1$, which is visually indistinguishable from $m=0.05$ for earlier times. Moreover, for later times, the maximum values of $E_{I,m}(t)$ and $E_{V,m}(t)$ are approximately $0.5\%$ and $1\%$, respectively, which are well within acceptable limits. Therefore, the subsequent sections of this article focus on the results obtained using $m=0.1$.

\begin{figure}[h!]
(a)\\
\begin{tikzpicture}
  \node at (0, 0) {\includegraphics[width=0.7\textwidth]{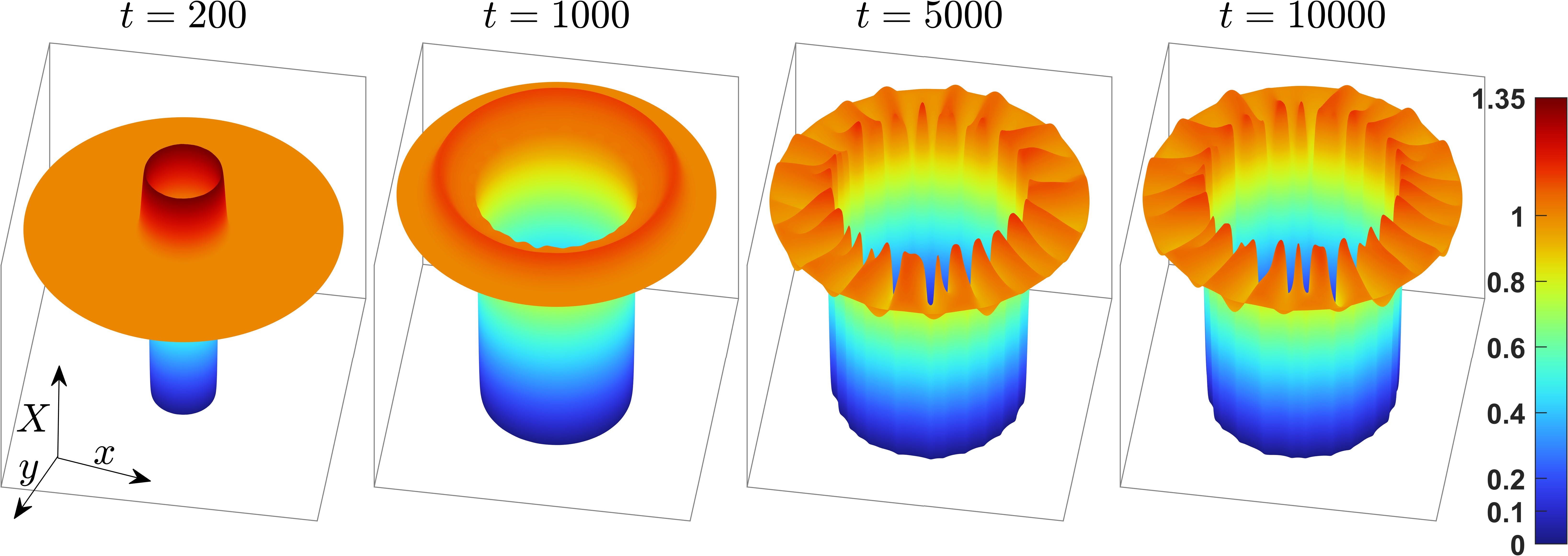}};
  \draw[->, thick] (-3.2, 1.5) -- (-2.75, 1.0); \node at (-2.95, 1.7) {Hump};
  \draw[->, thick] (-3.2, 1.5) -- (-4.6, 1.0) ;

  \draw[->, thick] (4.5, -1.6) -- (4.28, -0.2); \node at (4.45, -1.75) {Peaks};
  \draw[->, thick] (4.5, -1.6) -- (4.55, -0.1) ;

    \draw[->, thick] (3, -1.6) -- (3.6, -0.25); \node at (3, -1.8) {Valleys/};
    \node at (3, -2.1) {Sun-ray sources};
  \draw[->, thick] (3, -1.6) -- (3.3, -0.09) ;
\end{tikzpicture}

(b)\\
\begin{tikzpicture}
  \node at (0, 0) {\includegraphics[width=0.7\textwidth]{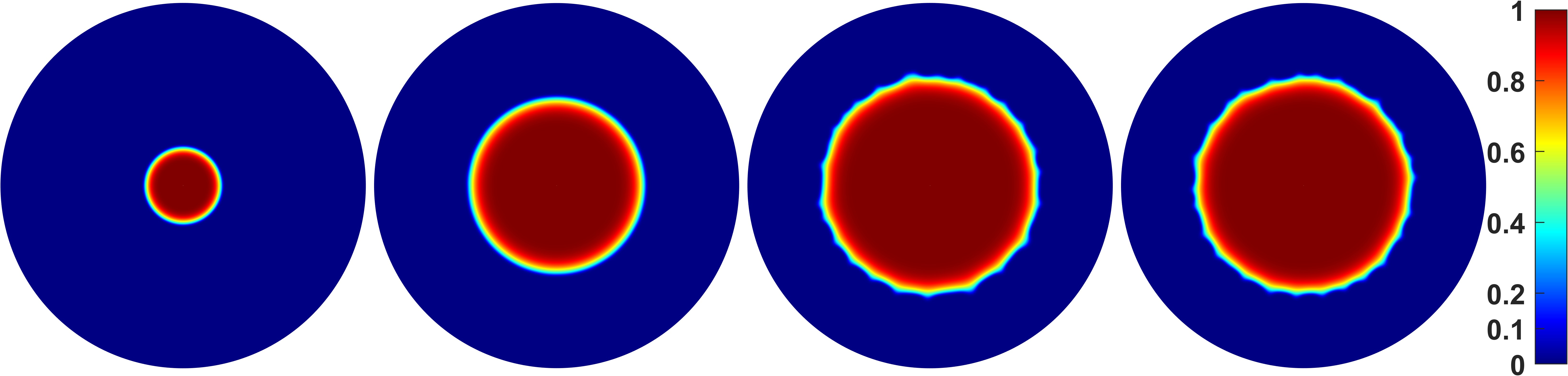}};
   \draw[->, thick] (3, 1.8) -- (3.88, 0.9); \node at (3, 2) {Sun-ray sources};
  \draw[->, thick] (3, 1.8) -- (3.38, 0.55) ;
  
\end{tikzpicture}
\vspace{-.3 cm}\caption{\label{surfXY_Qeffect} (a) Surface plots of the autocatalyst $X$ at different times for $Q=160,\delta=10$, (b) the spatio-temporal evolution of reactant $Y$ for the same parameter values.}
\end{figure}

\section{Non-linear simulations}
Figure \ref{surfXY_Qeffect}(a) displays the spatio-temporal evolution of the surface concentration of autocatalyst $X$, corresponding to Figure 1(a) in the main document. The red strips indicate humps in the surface, highlighted by arrows in the panels for $t=200$ and $t=1000$. These humps form due to the larger diffusion coefficient of the reactant and increased consumption by the autocatalyst, eventually spreading radially and diminishing in height. The unstable reaction front, with sun-ray patterns, always appears at smaller radial distances than these humps. In the panels for $t=5000$ and $t=10000$, arrows indicate peaks and valleys, where the valleys mark the sources of the sun-ray patterns seen in the autocatalyst $X$ surface. 

\begin{figure}[ht!]
    \centering
    \includegraphics[width=0.5\linewidth]{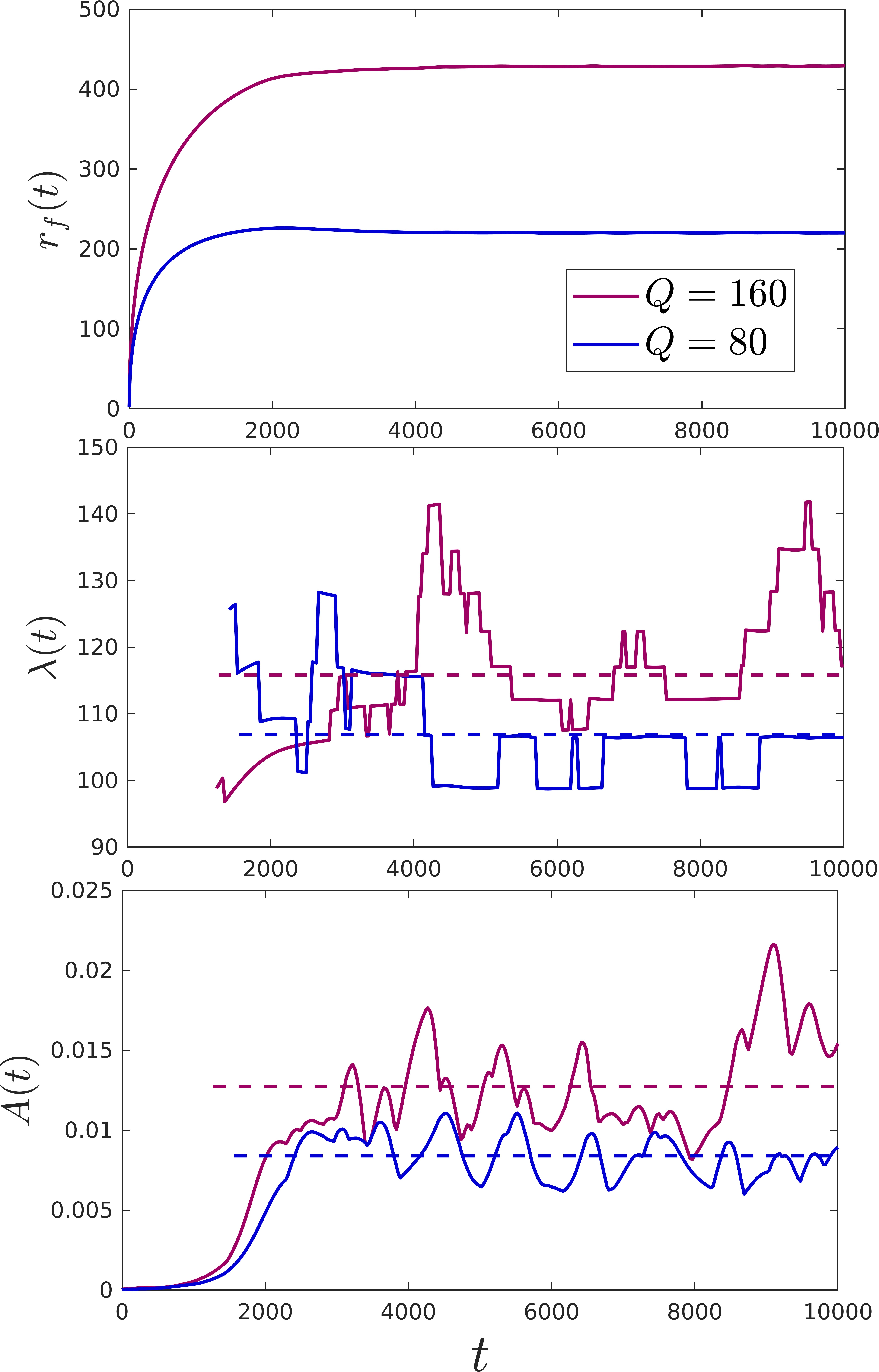}
     \begin{picture}(0,0)
    \put(-250,390){\makebox(0,0){(a)}}
    \put(-250,265){\makebox(0,0){(b)}}
    \put(-250,135){\makebox(0,0){(c)}}
\end{picture}
   \caption{
    (a) Temporal evolution of the reaction front position, \( r_f \), for \( Q = 80 \) and \( Q = 160 \) when \( \delta = 10 \). 
    (b) Temporal evolution of the wavelength, \( \lambda(t) \), for \( Q = 80 \) and \( Q = 160 \) when \( \delta = 10 \). Dashed lines indicate the mean wavelength, \( \lambda \). 
    (c) Temporal evolution of the amplitude, \( A(t) \), for \( Q = 80 \) and \( Q = 160 \) when \( \delta = 10 \). Dashed lines indicate the mean amplitude, \( A \).
    }
    \label{fig:temp_num}
\end{figure}

\section{Experimental set-up and data processing}

\begin{figure}[h!]\includegraphics[width=0.4 \textwidth]{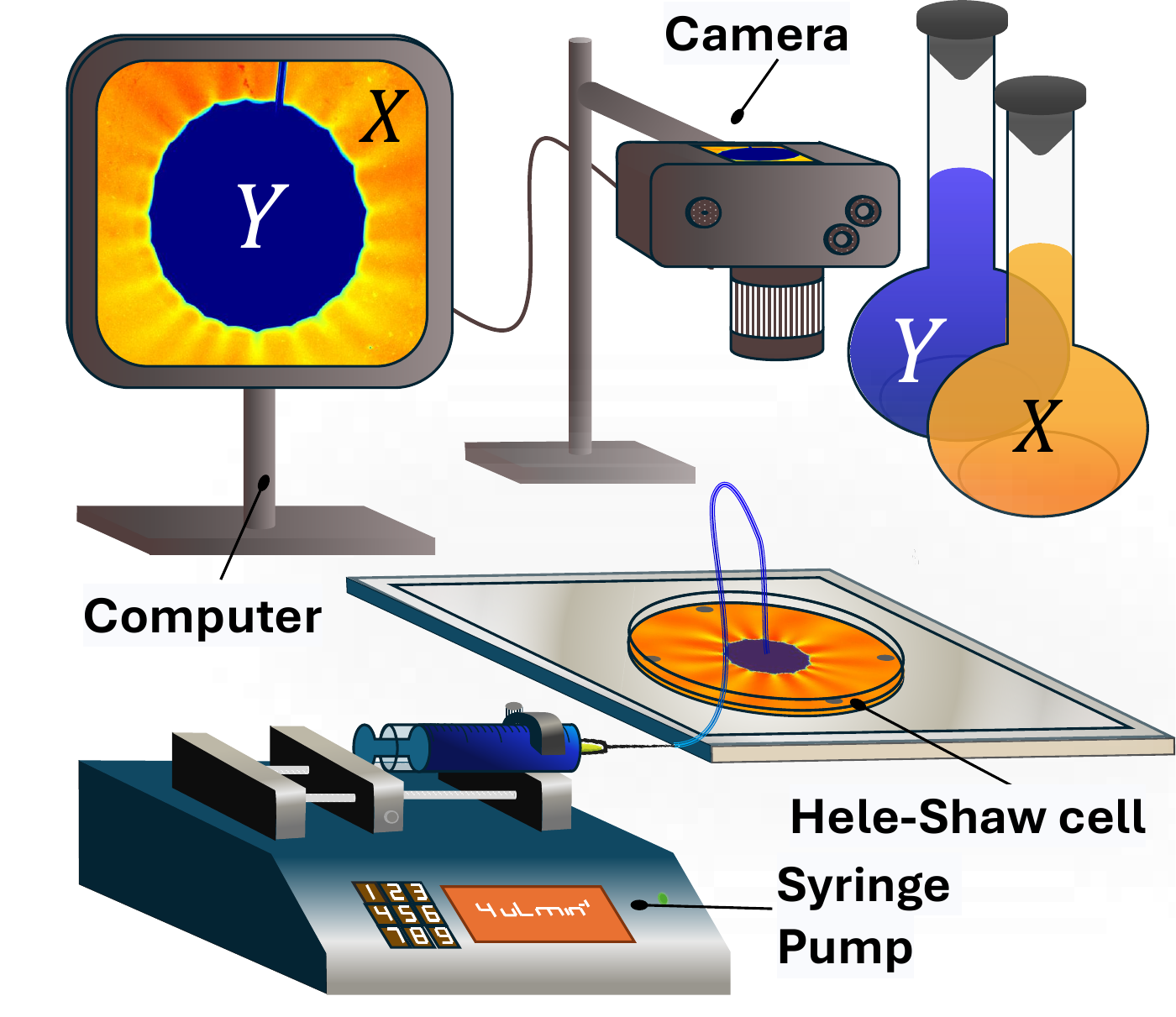}
\vspace{-.3 cm}\caption{\label{Fig: expset} Schematic of the experimental set-up.}
\end{figure} 

\begin{figure}[h!]\includegraphics[width=0.7 \textwidth]{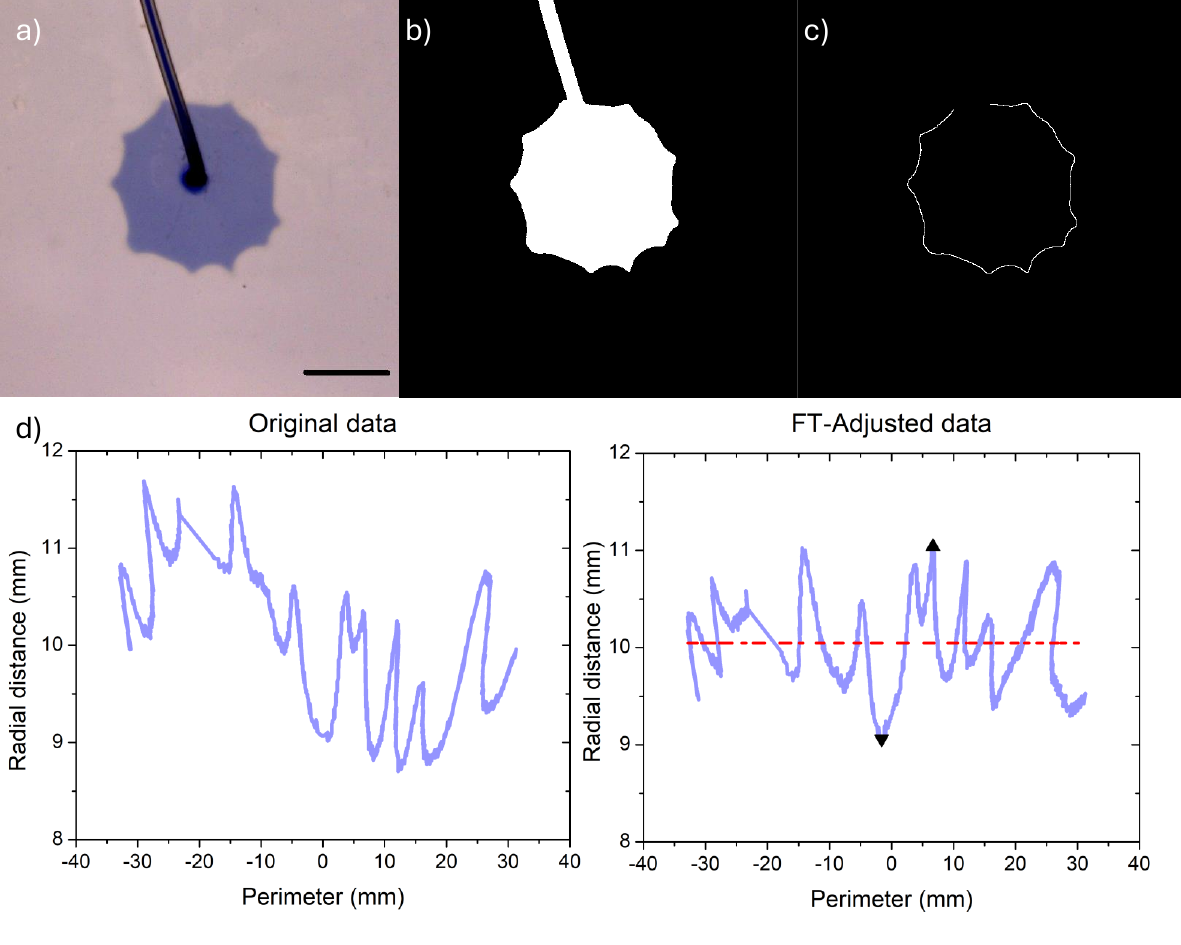}
\vspace{-.3 cm}\caption{\label{Fig: DataProcessing} (a) Original picture of the reactant solution Y, containing 30.6 mM of SPA, being continuously injected into the product solution X at a flow rate $\hat{Q}=1$ $\mu L$ $min^{-1}$, resulting in a stationary cellular front 10 mm away from the inlet. Scale bar: 10 mm. (b) Binarized image of the stationary front. (c) Outline of the front with the injection tube removed in the post-processing. (d) Comparison of the front position data in a radial coordinate set before and after FT-based filtering. The front's leading and trailing points are marked with upward and downward-facing triangles, respectively. The average front position is marked with a dashed red line.}
\end{figure}

Fig. S6 illustrates the experimental setup. The reactant Y is injected into the medium containing the autocatalyst X using a syringe pump, and the dynamics are recorded with a camera. Initially, digital images of the experiments are captured (Fig. \ref{Fig: DataProcessing}a). These images are binarized to enhance the contrast between the reacted and unreacted regions (Fig. \ref{Fig: DataProcessing}b). Following binarization, an outline delineating the reacted and unreacted regions is generated. In this step, the pixels along the outline are assigned a value of 255, while the remaining pixels are set to 0 (Fig. \ref{Fig: DataProcessing}c). The coordinates of these pixels are initially expressed in Cartesian coordinates. For further analysis, we transform them into radial coordinates ($r, \theta$), where $r$ is the radial distance of the point from the center of mass of the image, and $\theta$ is the angular position (Fig. \ref{Fig: DataProcessing}d). 

The experimentally obtained fronts are not perfectly circular due to the inevitable imperfections in flow. These slight deviations are corrected by applying a Fourier transform (FT) to the radial coordinates of the outline (Fig. \ref{Fig: DataProcessing}e). FT allows for filtering out the longest wavelength components in the data, corresponding to slight deviations from circularity. The process is demonstrated in Fig. \ref{Fig: DataProcessing}.
 
\section{Experimental results}

\begin{figure}
    \centering
    \includegraphics[width=0.5\linewidth]{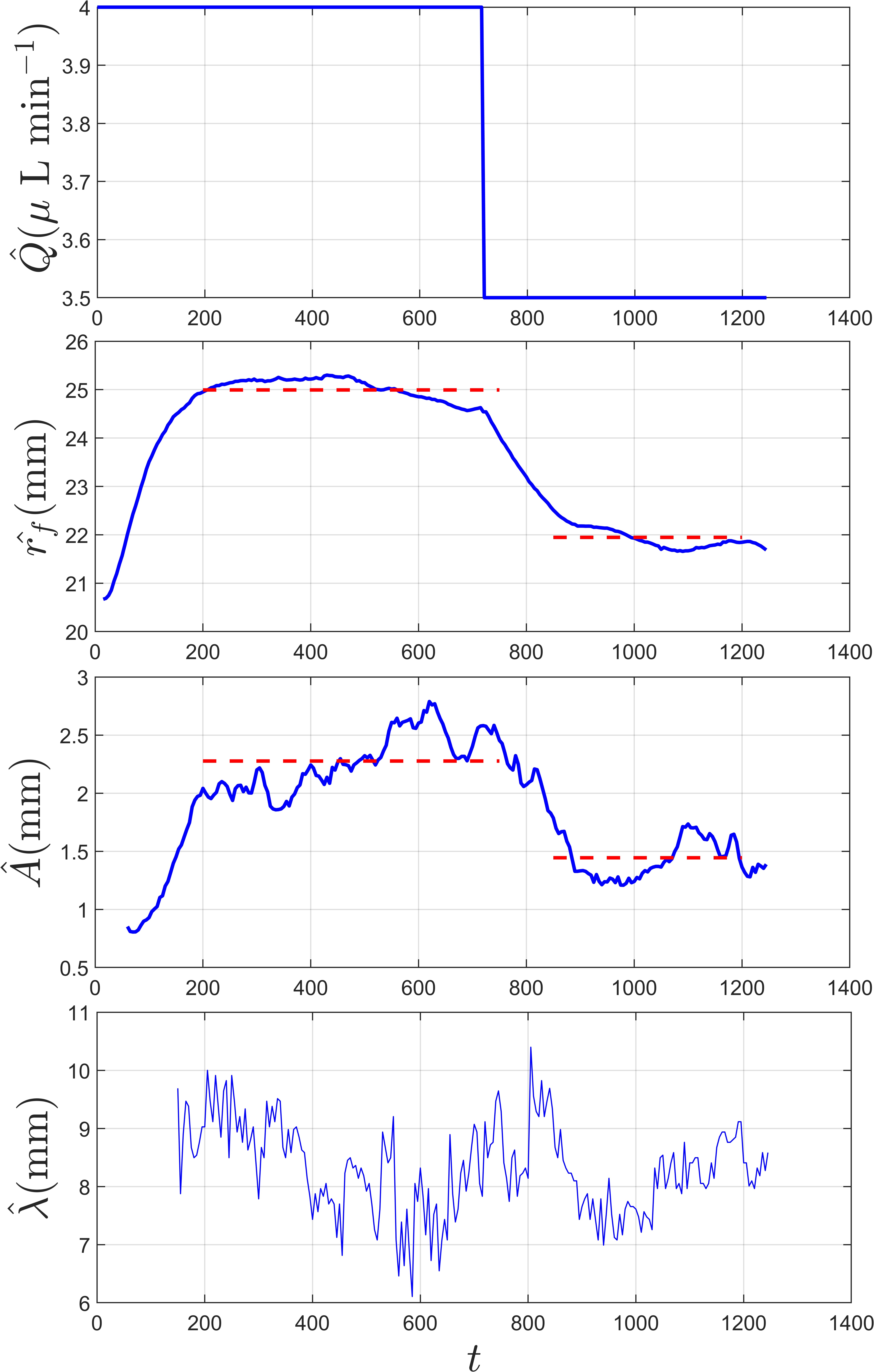}
         \begin{picture}(0,0)
    \put(-265,395){\makebox(0,0){(a)}}
    \put(-265,295){\makebox(0,0){(b)}}
    \put(-265,195){\makebox(0,0){(c)}}
    \put(-262,95){\makebox(0,0){(d)}}
\end{picture}
    \caption{Temporal evolution of key observables in response to changes in flow rate $\hat{Q}$: (a) imposed flow rate $\hat{Q}$ over time, and resulting variations in (b) front position, (c) amplitude, and (d) wavelength during the experiments. Full blue lines represent experimental data, while dashed red lines represent their average value. While the dependence of the stationary front position $\hat{r}_s$ and the cellular modulation amplitude $\hat{A}$ on the flow rate  $\hat{Q}$ are evident, there is no statistically significant relationship between the flow rate and the average modulation wavelength $\hat{\lambda}$.}
    \label{fig:experiments_timelapse}
\end{figure}

Upon setting the flow rate $\hat{Q}$ to a constant value (see Fig.~\ref{fig:experiments_timelapse}a), the front progresses until $\hat{v}_r = \hat{Q}/2\pi \hat{h} D_Xr $ equals the opposing RD velocity of the front $\hat{c}$. At this point, as seen in Fig.~\ref{fig:experiments_timelapse}b the front becomes stationary at the position $\hat{r}_s$.
Once the front stabilizes, the cellular modulation amplitude fluctuates around a steady value as cells merge and reappear. When the flow rate $\hat{Q}$ decreases, the front retracts closer to the inlet (Fig.~\ref{fig:experiments_timelapse}b), and the modulation amplitude (Fig.~\ref{fig:experiments_timelapse}c) adapts accordingly, eventually fluctuating around a new steady value.

Interestingly, Fig.~\ref{fig:experiments_timelapse}d shows that the average modulation wavelength remains constant, independent of the flow rate. This behavior is illustrated in Fig. \ref{fig:experiments_timelapse}, where a reduction in flow rate from 4 $\mu$L min$^{-1}$ to 3.5 $\mu$L min$^{-1}$ causes the amplitude to decrease from 2.1 mm to 1.5 mm.

\newpage

\bibliography{Main}